\documentclass{aa}  
\usepackage{placeins}
\usepackage{graphicx}
\usepackage{caption}
\usepackage{txfonts}
\usepackage{natbib}
\bibpunct{(}{)}{;}{a}{}{,} 
\usepackage[]{hyperref}
\hypersetup{colorlinks,
citecolor=blue
}
\begin{document} 

\title{A systematic search for dormant galaxies at z $\sim$ 5$-$7 from the JWST NIRSpec archive}
\titlerunning{Systematic search for dormant galaxies at z $\sim$ 5$-$7}

   \author{Alba Covelo-Paz\inst{1},
          Corentin Meuwly\inst{1},
          Pascal A. Oesch\inst{1,2,3},
          Callum Witten\inst{1},
          Andrea Weibel\inst{1},
          Cristian Carvajal-Bohorquez\inst{4},
          Laure Ciesla\inst{4},
          Emma Giovinazzo\inst{1},
          Gabriel Brammer\inst{2,3}
          }

   \institute{Department of Astronomy, University of Geneva, Chemin Pegasi 51, 1290 Versoix, Switzerland\\ 
   \email{alba.covelopaz@unige.ch}
   \and
   Cosmic Dawn Center (DAWN), Copenhagen, Denmark
   \and
   Niels Bohr Insitute, University of Copenhagen, Jagtvej 128, 2200 Copenhagen, Denmark
   \and 
   Aix Marseille Univ, CNRS, CNES, LAM, Marseille, France
}
\authorrunning{A. Covelo-Paz et al.}
   \date{Received XX; accepted YY}

 
\abstract{JWST has revealed a population of ``dormant'' galaxies at $z>5$ that have recently halted their star formation and are characterized by weak emission lines and significant Balmer breaks. Until now, only four such galaxies have been reported at $z>5$, three with low stellar masses, $M_*<10^9M_\odot$ (so-called mini-quenched galaxies), and one massive quiescent galaxy with $M_*=10^{10.2}M_\odot$; no such galaxy had been reported at intermediate masses.  
Here, we present a systematic search for dormant galaxies at $5<z<7.4$ that halted star formation at least 10 Myr before the time of observation. To do this, we made use of all the publicly available NIRSpec prism data in the DAWN JWST Archive (DJA) and select galaxies with low H$\alpha$ equivalent widths ($EW_{0}<50$\AA) and strong Balmer breaks ($F_{\nu,4200}/F_{\nu,3500}>1.4$). We find 14 dormant galaxies with stellar masses ranging from $10^{7.6}-10^{10.5}$, revealing an intermediate-mass population. By construction, these 14 sources are located about 1 dex below the star-forming main sequence. Their star formation histories suggest that they halted star formation between 10 and 25 Myr before the time of observation which, according to models, is comparable with the timescales of internally regulated bursts driving a ``breathing'' mode of star formation. 
Our results show that $\sim1\%$ of the galaxies in the DJA are in a dormant phase of their star formation histories, and they span a wide stellar mass range. These galaxies can be empirically selected using only their spectral features in NIRSpec prism data.}


   \maketitle
%
\section{Introduction}

During the evolution of a galaxy, its star formation rate (SFR) can vary greatly, 
reflecting significant changes in its environment, such as interactions with other galaxies or inflows of cold gas, or internal processes such as feedback from supernovae or active galactic nuclei (AGNs)
\citep{man&belli18}. 
At $z \gtrsim 5$, the star formation histories (SFHs) of low-mass galaxies are often characterized as being stochastic: low-mass systems have relatively shallow gravitational potentials, making them very sensitive to their environment, galactic interactions, dynamics in dense environments \citep{peng+10,Bluck+20,Alberts+24}, and stellar feedback \citep{Hopkins23,Gelli25}. These effects can strongly influence galaxies' star formation and hence can lead a galaxy to rapidly transition from a starburst to a quenched state. 
This ``burstiness'' can be observed in the dispersion of the main sequence \citep[e.g.,][]{Looser+23,Dome24,Dressler+24,clarke+24,Ciesla24} as well as in simulations \citep[e.g.,][]{fukuda00,immeli04,bournaud07,Elmegreen09,fg18,Tacchella20}.
Bursty SFHs imply that galaxies can temporarily enter a ``dormant'' or ``lulling'' phase, which is followed by an increase in their SFRs. Such dormant galaxies at $z\gtrsim5$ have long been predicted by simulations \citep[e.g.,][]{Wyithe14} and can now finally be observed with JWST \citep{Gardner+06,Rigby+23}. Constraints on the frequency and duration of bursting and quenching phases are currently insufficient to inform models of galaxy evolution. The key to building robust and accurate models of the life cycle of early galaxies is observational constraints on their different phases, which is only possible with the identification of galaxies across all of the stages of their SFHs.

The H$\alpha$ emission line has been extensively used for determining the SFRs of galaxies over a wide range of redshifts.
This line results from the $3\rightarrow2$ recombination of hydrogen gas after its ionization by high-energy photons. The only stars that emit such energetic photons are the young, massive stars, which are also the most short-lived. Therefore, H$\alpha$ serves as a tracer to detect recent star formation in a galaxy, typically over a period of $\lesssim$ 10 Myr prior to observation \citep[e.g.,][]{kennicutt+98,McClymont25}.
Until recently, SFR determination relied extensively on the H$\alpha$ line at $z < 3$ \citep[e.g.,][]{Erb+06,Geach+08,Forster09,Hayes10,Sobral+13} 
and could only be extended up to $z \lesssim 4.5$ by relying on broadband \emph{Spitzer} photometry \citep[e.g.,][]{Bollo+23}.
In this respect, JWST has lifted the veil at high redshifts, offering new insights and allowing us to detect the H$\alpha$ line with spectroscopy at $z>3$ \citep[e.g.,][]{Matharu24,Nelson+24,CoveloPaz24,Zhang25} and with narrowband \citep[e.g.,][]{Pirie24} and medium-band \citep[e.g.,][]{simmonds24} photometry up to $z \lesssim 7.4$.
The access to this line is of particular importance in the context of dormant galaxies, since they are expected to show weak H$\alpha$ emission due to their low recent SFR.

JWST NIRSpec prism spectroscopy has uncovered a handful of low-mass ($M_*<10^9M_\odot$) dormant galaxies, the so-called mini-quenched galaxies, at $z>5$ \citep{Strait+23,Looser+24,Baker+25}. These objects show strong Balmer breaks and weak emission lines, tracing a lack of recent star formation; however, their UV luminosities are strong, pointing toward increased star formation on longer timescales ($\sim100$ Myr). The picture is different at higher masses, where only the spectrum of one massive ($M_*=10^{10.23}M_\odot$) quenched galaxy has been reported at $z>5$ so far \citep{Weibel+24}; it ceased star formation 50–100 Myr before observation and is likely to remain quiescent. No dormant galaxies have been reported spectroscopically in this redshift range with intermediate masses ($M\sim10^9-10^{10}M_\odot$), limiting our understanding of burstiness and quenching at this mass range. 

Efforts to systematically search for high-redshift dormant galaxies have been made using photometry: \citealt{Alberts+24} found 12 dormant candidates at $3<z<7$ via UVJ selection with NIRCam and MIRI imaging, and \citealt{Trussler25} report 16 ``smouldering'' candidates at $5<z<8$ detected via color selection with NIRCam imaging.
Moreover, \citealt{Baker25c} found 745 massive quiescent galaxies at $2<z<7$ via UVJ selection with NIRCam imaging. 
We sought to complement these approaches with JWST spectroscopic data. Thanks to the low spectral resolution and high sensitivity of NIRSpec prism data, we can simultaneously constrain the rest-optical continuum (i.e., the Balmer break) and the H$\alpha$ emission line up to $z<7.4$. Thus, we utilized constraints on the H$\alpha$ line and Balmer break strengths, as tracers of recent quenching, to perform a systematic search for dormant galaxies in all the publicly available NIRSpec prism data. 

This paper is structured as follows: in Section \ref{sec:data}, we discuss the methods used to build the galaxy catalog and measure the spectral features. 
In Section \ref{sec:dorm_select}, we describe the selection of dormant galaxies. 
In Section \ref{sec:res}, we discuss the results, before concluding in Section \ref{sec:conclu}. 
Throughout this work, we adopt a concordance flat $\Lambda$CDM cosmology with $H_0=70\,\textrm{km}\,\textrm{s}^{-1}\,\textrm{Mpc}^{-1}$, $\Omega_m=0.3$, and $\Omega_\Lambda=0.7$. Magnitudes are listed in the AB system \citep{Oke&Gunn}. 

\section{Data}
\label{sec:data}
\subsection{Observations and data selection}

We retrieved our data from the fourth version of the DAWN JWST Archive (DJA)\footnote{\url{https://s3.amazonaws.com/msaexp-nirspec/extractions/public_prelim_v4.2.html}}, which compiles $\sim67\,000$ public JWST NIRSpec observations collected by various public programs, reduced following the process described in \citet{heintz24} for version 2, \citet{degraaf25} for version 3, \citet{Valentino25} for version 4, and \citet{Pollock25} for version 4.2. 
As we aim to characterize the H$\alpha$ emission of high-redshift galaxies, we searched for all prism spectra covering the rest-frame $6\,565\AA$ region. Since NIRSpec has an (extended) wavelength coverage up to 5.5 $\mu$m (see e.g., \citealt{Pollock25}), the detection of the H$\alpha$ line is restricted to $z\lesssim7.4$.
To guarantee an accurate spectroscopic redshift determination of our sources, we only considered galaxies with $z>5$, which show a Ly-$\alpha$ break that we can use to independently confirm the redshift measurement. On top of that, we made use of the ``grade" assigned to each spectrum by visual inspection, included in the DJA. Only galaxies with grade 3 are considered to have robust redshift measurements, and thus we selected only galaxies with this grade.
Finally, we removed 82 broad-line AGNs from the sample reported in \citet{Kocevski+24},
as their broadened H$\alpha$ line does not directly probe star formation. After applying these filters, our final galaxy catalog is composed of 1,598 objects with NIRSpec prism data.

In addition to spectroscopy, we used photometric measurements from JWST/NIRCam as well as \textit{Hubble} Space Telescope (HST) imaging, where available. We followed the procedure outlined in \citet{Weibel+24} to derive point spread function (PSF)-matched, aperture-corrected photometry based on the imaging data that are also available in the DJA. Note that most sources in our sample do have both NIRSpec prism spectra and multi-wavelength photometry, with only $\sim15\%$ of our sample not having any photometry at all. The spectra lacking photometry were calibrated on the DJA using the standard JWST pipeline, without extra slit-loss correction.

\subsection{SED fitting}
To determine the galaxy properties, we fit the spectroscopy and photometry of our sources simultaneously using the \texttt{Bagpipes} spectral energy distribution (SED) fitting code \citep{Bagpipes}. We used a first-order polynomial to derive residual slit-loss corrections of the spectrum to the photometry. 
A few galaxies in our catalog are found in lensed fields, and we accounted for this by dividing the fluxes of both the spectra and the photometry by the magnification obtained from the lensing models of MACS-J0647 \citep{Meena23} and ABELL2744 \citep{Furtak23,Price25}.

For all sources, the redshift was fixed to the spectroscopic redshift in the DJA, derived from \texttt{msaexp} \citep{msaexp}. We used the BPASS stellar population synthesis models and adopted a nonparametric SFH following the continuity prior from \citet{Leja+19} with age bins set at 0, 3, 10, 25, 50, 100, 250, 500, and 700 Myr. 
We used a \citet{Salim+18} dust attenuation curve, where we allowed the $2\,175 \AA$ bump strength to vary from $B=0-0.5$ following the usual values at $5<z<7$ reported by \citet{Markov25} and allowed the deviation from the \citet{Calzetti+00} slope to vary from $\delta=-0.8-0.3$. We allowed the visual attenuation to vary from $A_V=0-5$ mag, and the metallicity from $Z=0.1Z_\odot-Z_\odot$, using uniform priors. Based on the resulting V-band attenuation, $A_V$, we further derived the dust attenuation at the  H$\alpha$ wavelength, $A_{\textrm{H}\alpha}$, based on the \citet{Salim+18} curve, assuming that the nebular reddening is the same as the continuum reddening. 

\subsection{Spectral features}
We first extracted the flux of the H$\alpha$ line by fitting the lines and underlying continuum simultaneously with three Gaussian profiles (one for the H$\alpha$ line at $6\,564.59\AA$, and one for each line in the [NII] doublet at $6\,549.84\AA$ and  $6\,585.23\AA$) and a constant component for the underlying continuum. 
Based on this, we derived the observed H$\alpha$ luminosities of our sample of galaxies and computed the intrinsic luminosities by accounting for dust attenuation:
\begin{equation}
    \label{eq:att}
    L_{\text{H}\alpha\text{,int}} = L_{\text{H}\alpha\text{,obs}} \cdot 10^{0.4A_{\text{H}\alpha}}.
\end{equation}

\begin{figure*}
    \centering
    \includegraphics[width=0.85\linewidth]{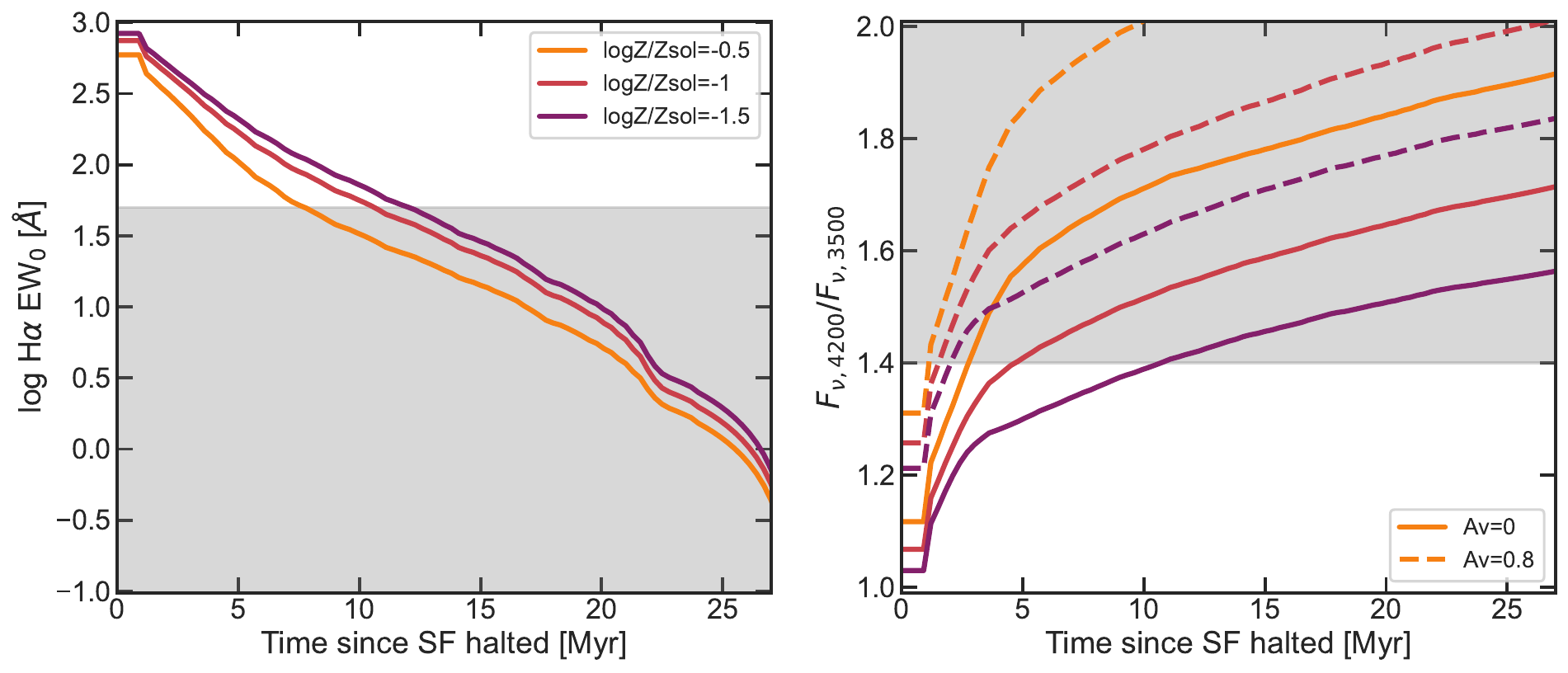}     
    \caption{Balmer beaks and rest-frame H$\alpha$ EWs of model galaxies that halted star formation at a grid of times before observation. The \texttt{Bagpipes} models assume a constant SFH of 100 Myr before star formation is stopped. They were computed with three different metallicities and two dust scenarios: $A_V=0$ and $A_V=0.8$. The gray region denotes the parameter space from which galaxies were selected for our sample. We find that the evolution of the Balmer breaks is metallicity- and dust-dependent, while H$\alpha$ EWs experience minimal variations between different metallicities, with no variations with different dust attenuation since we applied the same dust screen to the nebular lines as to the continuum.}
    \label{fig:model_halted}
     \end{figure*}

We computed the recent SFR from the H$\alpha$ emission as:
\begin{equation}
    \label{eq:SFR}
    \text{SFR}_{\text{H}\alpha} = 10^{-41.67}\frac{L_{\textrm{H}\alpha,\textrm{int}}}{\textrm{erg/s}} ,
\end{equation}
where the conversion factor has been derived by \citet{Shapley+23} from low-metallicity BPASS population models and assuming an upper-mass initial mass function (IMF) cutoff of $100M_\odot$.

We measured the H$\alpha$ equivalent width (EW) by dividing the H$\alpha$ flux by the H$\alpha$ continuum, which we obtained from the \texttt{Bagpipes} fits instead of the spectra because the low S/N in certain sources did not allow for a good fit to the continuum. We then divided this EW by $(1+z)$ to obtain the rest-frame H$\alpha$ EW. Finally, we measured the strength of the Balmer break as the ratio of the $f_\nu$ values between a red window from $3400-3600\AA$ and a blue window from $4150-4250\AA$. As the uncertainties of this ratio are considerably high when measured from the observed spectrum for many galaxies in our sample, we computed the Balmer break as the ratio between the red and blue windows from the \texttt{Bagpipes} fits.  
     
\section{Selection of dormant galaxies}
\label{sec:dorm_select}

To find an appropriate selection of dormant galaxies, we started by computing \texttt{Bagpipes} model SEDs to determine expected H$\alpha$ EW values for galaxies that have recently halted star formation. We adopted a constant SFH for 100 Myr, before the SFR is switched off, and assumed no dust attenuation. We computed the model over a grid of metallicities and time since star formation halted, and computed the rest-frame H$\alpha$ EW and Balmer break. The results are presented in Fig. \ref{fig:model_halted} and show that, for galaxies that halted star formation at least 10 Myr before the time of observation, the H$\alpha$ EW is consistently below $\sim50\AA$, and we note that this does not change in the presence of dust.  
Thus, a galaxy that is still undergoing star formation would not have a rest-frame EW as low as $<50\AA$, under the assumption that the escape fraction of ionizing photons is zero (see \citealt{Emma+25} for further discussion of potential degeneracies) and assuming that there is no strong differential extinction between the continuum and emission line.
However, if the faint Balmer lines were due to an increased Lyman continuum escape fraction instead of quenched star formation, we would not expect the formation of Balmer breaks in the spectra and we would expect very steep UV slopes ($\beta_{\rm UV}$).

Another characteristic feature of quenched galaxies is the formation of a Balmer break. 
Fig. \ref{fig:model_halted} shows that the Balmer break of galaxies that halted star formation at least 10 Myr before the time of observation is consistently above 1.4 for dust-free galaxies, and above 1.6 for our limiting case, $A_V=0.8$. This lies above the average values of $F_{\nu,4200}/F_{\nu,3500}\lesssim1.1$ in the $5<z<7$ galaxies reported by \citet{Roberts-Borsani24} (see also \citet{Kuruvanthodi24} for $z>7$ objects). We selected the sources from our sample with a $F_{\nu,4200}/F_{\nu,3500}>1.4$ and removed dusty galaxies with $A_V>0.8$ from the selection, as these tend to show large $F_{\nu,4200}/F_{\nu,3500}$ ratios that are not caused by Balmer breaks.
From this set, we selected the sources with a rest-frame H$\alpha$ EW $<50\AA$ and $S/N>2$ in the continuum around the H$\alpha$ line to ensure reliable EW measurements. This yielded a sample of 14 potentially dormant galaxies, which are presented in Table \ref{tab:dormants}. We note that in this selection there are only two galaxies with a Balmer break $<1.6$, both with low dust attenuation. The only galaxy reaching a value close to $A_V=0.8$ has a Balmer break of $\sim1.9$, meaning that all our selected galaxies fulfill the criterion of quenching $>10$ Myr before observation according to Fig. \ref{fig:model_halted}.

\section{Results}
\label{sec:res}

\subsection{Sample}

\begin{figure*}[h!]
\includegraphics[width=\linewidth]{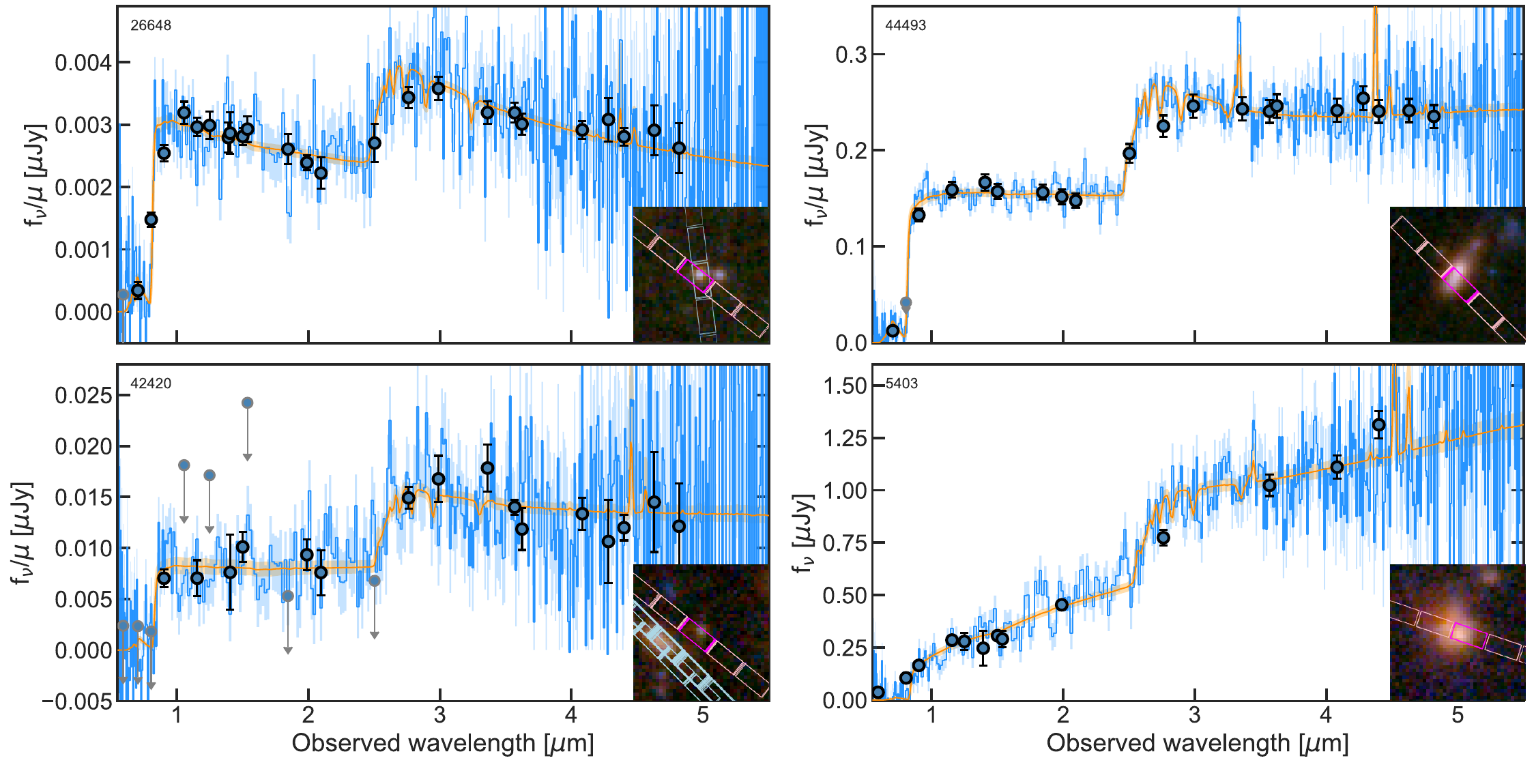}
    \caption{Observed $f_\nu$ spectra (blue) and posterior spectra from \texttt{Bagpipes} (orange) for four of the dormant galaxies selected in this work (the full set of spectra is shown in the appendix). The filled circles are the photometry values, which we fit simultaneously with the spectra. We observe strong Balmer breaks and weak emission lines. Insets: are the red-green-blue (RGB) cutouts of the galaxies, with the slits used for these observations overplotted in pink, showing the large size range of these sources.  } 
    \label{fig:spectra_sample}
\end{figure*}

The spectra of the 14 dormant galaxies selected in this work were obtained from various programs: JADES (GTO-1180, GO-3215, \citealt{jades,udeep}), GTO WIDE (GTO-1214, \citealt{gto_wide}), GO 1433 \citep{MACS-J0647}, UNCOVER (GO-2561, \citealt{uncover}), and CAPERS (GO-6368; PI: Dickinson). A subset of the spectra of these sources is shown in Fig. \ref{fig:spectra_sample}, with the full sample shown in the appendix in Fig. \ref{fig:spectra}. As can be seen in these figures, all the sources except one have photometry that is consistent with the spectra, with one source lying in an area without photometry available. These 14 dormant galaxies have redshifts in the range of $5.04<z<6.78$ and show varied spectral features: certain galaxies remain with weak but clear emission lines, while others show no emission lines and hints of what could be absorption lines, although more exposure time would be needed to confirm this. From the red-green-blue (RGB) stamps of these sources, we find a variety of morphologies: some are compact and others are more extended. Several galaxies in the sample have close companions, which opens the possibility that galaxy-galaxy interactions influence the recent quenching of the targeted sources. 

The UV slopes of this sample are typically steep, $\beta_\textrm{UV}\sim-2$, with the exception of the most massive galaxy in our sample, $M_*=10^{10.47}M_\odot$, with a $\beta_\textrm{UV}=-0.90\pm0.07$ and an extraordinarily strong dust attenuation of $A_V=0.74_{-0.12}^{+0.14}$, which is close to the limit of $A_V<0.8$ that we imposed and an outlier compared to the median of our sample of dormants, $A_{V,\textrm{med}}=0.14$. Despite this strong dust attenuation, there is unambiguously a Balmer break on its spectrum, and the SFH of this galaxy shows that it halted star formation 10 Myr before the time of observation, explaining the formation of the Balmer break. As introduced in Section \ref{sec:dorm_select}, these galaxies are unlikely to experience a high Lyman continuum $f_{\rm esc}$, as the typical $\beta_\textrm{UV}$ expected for such galaxies would be steeper ($\beta_\textrm{UV}<-2.5$, see \citealt{Emma+25}). This supports the idea that the galaxies in our sample are indeed in an off mode. 

\begin{figure}[]
    \centering  
    \includegraphics[width=\columnwidth]{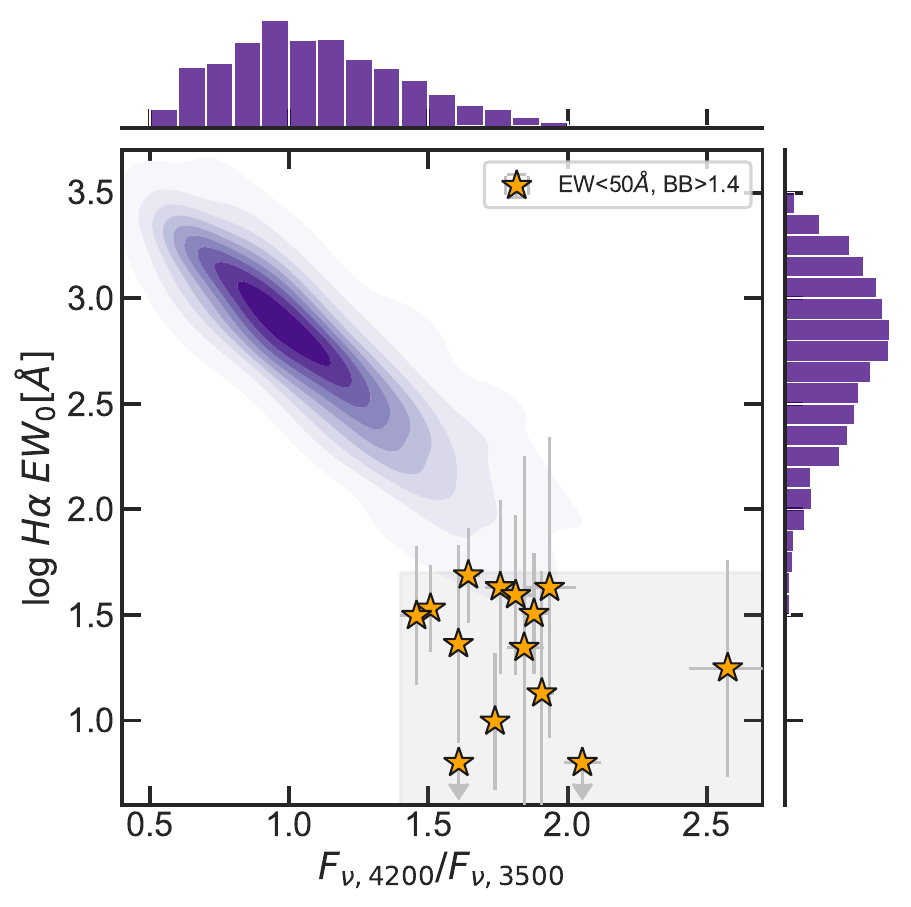}    
    \caption{Distribution of Balmer break ratios and rest-frame H$\alpha$ EWs of our 14 dormant galaxies, highlighted against the distribution of the total galaxy sample. We find a strong correlation between the two quantities in the entire galaxy population at $5<z<7.4$, which is a key result for our understanding of burstiness. The dormant galaxies selected in this work lie on the edge of this distribution (shadowed in gray), with the strongest Balmer breaks and the weakest EWs.}
    \label{fig:histograms}
     \end{figure}

\begin{table*}
\caption{Properties of the dormant galaxies selected in this work, sorted by redshift}
    \centering
\renewcommand{\arraystretch}{1.4}
    \begin{tabular}{cccccccccccc}
    \hline
     source & PID$^*$ & RA & Dec & $z_\textrm{spec}$ & H$\alpha$ EW & BB & $\log M_*$ & $\beta_\textrm{UV}$ & $\textrm{SFR}_{10}/\textrm{SFR}_{100}$ & $t_\textrm{mw}$$^\dagger$\\
     ID &  & deg & deg &&  $\AA$ & & $M_\odot$ & &&Myr\\
     \hline
68913 & 1181 & 189.210734 & 62.147085 & 5.04 & $39\pm15$ & $1.81_{-0.06}^{+0.06}$ & $9.28_{-0.04}^{+0.05}$ & $-2.23_{-0.04}^{+0.07}$ & $0.131_{-0.067}^{+0.082}$ & $262_{-58}^{+45}$\\
9483 & 6368 & 215.013250 & 52.974686 & 5.08 & $32\pm9$ & $1.88_{-0.04}^{+0.05}$ & $9.27_{-0.05}^{+0.04}$ & $-1.80_{-0.06}^{+0.07}$ & $0.144_{-0.052}^{+0.067}$ & $266_{-43}^{+40}$\\
7395 & 1214 & 150.136830 & 2.251830 & 5.11 & $49\pm11$ & $1.64_{-0.03}^{+0.04}$ & $9.79_{-0.05}^{+0.03}$ & $-1.70_{-0.06}^{+0.05}$ & $0.198_{-0.058}^{+0.085}$ & $209_{-44}^{+42}$\\
4188 & 1433 & 101.908377 & 70.251585 & 5.26 & $18\pm9$ & $2.57_{-0.14}^{+0.18}$ & $7.67_{-0.15}^{+0.15}$ & $-1.56_{-0.20}^{+0.28}$ & $0.106_{-0.106}^{+0.269}$ & $411_{-55}^{+50}$\\
11237 & 6368 & 214.947354 & 52.918977 & 5.49 & $-9\pm11$ & $1.61_{-0.03}^{+0.03}$ & $8.93_{-0.02}^{+0.04}$ & $-2.37_{-0.03}^{+0.02}$ & $0.001_{-0.001}^{+0.009}$ & $134_{-11}^{+17}$\\
16756 & 6368 & 214.879629 & 52.835771 & 5.66 & $-5\pm11$ & $2.05_{-0.07}^{+0.07}$ & $9.21_{-0.05}^{+0.04}$ & $-2.10_{-0.05}^{+0.06}$ & $0.010_{-0.010}^{+0.029}$ & $288_{-48}^{+50}$\\
26648 & 2561 & 3.591447 & -30.396666 & 5.66 & $23\pm11$ & $1.61_{-0.03}^{+0.03}$ & $7.59_{-0.03}^{+0.03}$ & $-2.27_{-0.04}^{+0.04}$ & $0.003_{-0.003}^{+0.008}$ & $81_{-15}^{+34}$\\
44493 & 2561 & 3.593161 & -30.346475 & 5.67 & $10\pm3$ & $1.74_{-0.02}^{+0.02}$ & $9.56_{-0.04}^{+0.04}$ & $-1.96_{-0.03}^{+0.03}$ & $0.021_{-0.017}^{+0.016}$ & $176_{-28}^{+42}$\\
42420 & 2561 & 3.595939 & -30.371346 & 5.79 & $43\pm30$ & $1.94_{-0.09}^{+0.09}$ & $8.33_{-0.04}^{+0.04}$ & $-2.08_{-0.08}^{+0.11}$ & $0.028_{-0.027}^{+0.045}$ & $245_{-61}^{+62}$\\
5403 & 1214 & 150.109947 & 2.259070 & 5.88 & $13\pm8$ & $1.91_{-0.04}^{+0.04}$ & $10.48_{-0.05}^{+0.05}$ & $-0.91_{-0.07}^{+0.08}$ & $0.002_{-0.002}^{+0.006}$ & $130_{-20}^{+27}$\\
27376 & 1181 & 189.182645 & 62.246510 & 5.91 & $34\pm7$ & $1.51_{-0.03}^{+0.03}$ & $9.22_{-0.05}^{+0.04}$ & $-1.80_{-0.04}^{+0.05}$ & $0.001_{-0.001}^{+0.004}$ & $67_{-12}^{+30}$\\
201850 & 3215 & 53.148424 & -27.801749 & 6.07 & $22\pm20$ & $1.84_{-0.06}^{+0.07}$ & $8.46_{-0.03}^{+0.04}$ & $-1.92_{-0.09}^{+0.09}$ & $0.007_{-0.007}^{+0.021}$ & $215_{-59}^{+51}$\\
235403 & 6368 & 214.878295 & 52.829096 & 6.54 & $31\pm11$ & $1.46_{-0.06}^{+0.07}$ & $9.17_{-0.09}^{+0.10}$ & $-2.38_{-0.07}^{+0.08}$ & $0.014_{-0.014}^{+0.042}$ & $150_{-43}^{+55}$\\
61471 & 2561 & 3.570125 & -30.336790 & 6.78 & $43\pm18$ & $1.76_{-0.05}^{+0.05}$ & $8.90_{-0.04}^{+0.04}$ & $-1.78_{-0.09}^{+0.09}$ & $0.012_{-0.012}^{+0.030}$ & $190_{-48}^{+55}$\\    \hline
    \end{tabular}
\label{tab:dormants}
\vspace{0.5em}

\begin{minipage}{\textwidth}
$^*$\small The program IDs correspond to the following surveys. 1181: JADES \citep{jades}. 1214: GTO WIDE \citep{gto_wide}. 1433: MACS-J0647 \citep{MACS-J0647}. 2561: UNCOVER \citep{uncover}. 3215: JADES Ultra-Deep \citep{udeep}. 6368: CAPERS. \\
$^\dagger$\small mass-weighted age
\end{minipage}
\end{table*}

\begin{figure*}
\centering
\includegraphics[width=0.8\linewidth]{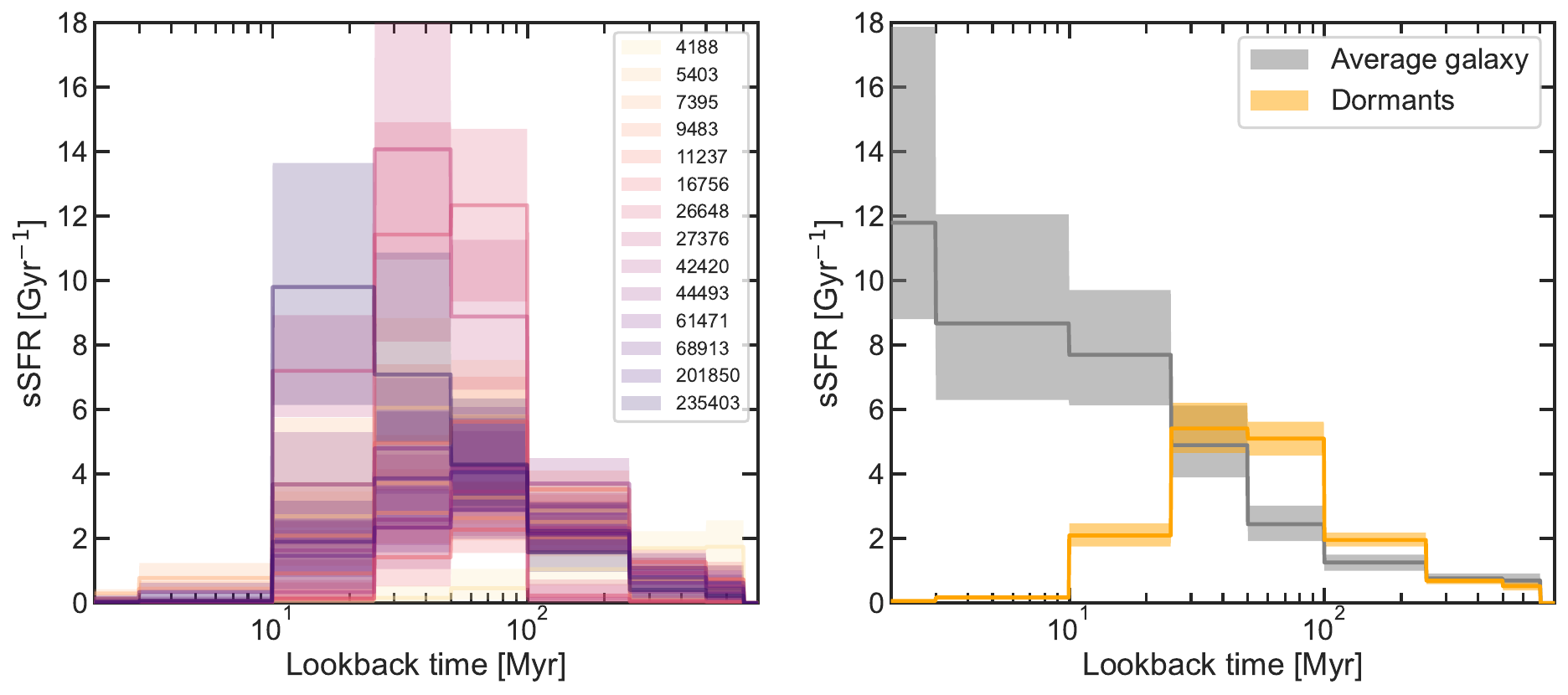}
    \caption{Past SFHs of the dormant galaxies in our sample. Left: sSFRs vs lookback time for all 14 individual sources. Shaded areas correspond to the 16-84th percentiles of the bagpipes posterior SFHs.
    Right: Mean sSFH of the 14 dormant galaxies against that of the entire sample of $1,598$ objects. The time bins for the continuity SFH were set to be at 0, 3, 10, 25, 50, 100, 250, 500, and 700 Myr. We observe that star formation halted in these galaxies between 10 and 25 Myr before observation. Previous to that, all galaxies experienced a continuous rise in star formation before it halted, except the galaxy with ID 4188, which experienced this burst 250 Myr before observation and whose star formation has been gradually decreasing ever since.}
    \label{fig:sfhs}
\end{figure*} 

Our dormant galaxies show Balmer breaks that span from $1.5-2.6$, which is much higher than the median value of the full population in the DJA, $F_{\nu,4200}/F_{\nu,3500}=1.1$. The rest-frame EWs of our sample span from $-9-50\AA$, much below the median $580\AA$ of the DJA catalog. 
Fig. \ref{fig:histograms} shows our selection of galaxies within the Balmer break and EW distributions of the original catalog. We find a
strong correlation between the two quantities in the entire galaxy population at $5<z<7.4$. The scatter of this distribution is a result of different SFHs of the full galaxy population \citep[e.g.,][]{Mintz+25} and will be analyzed in more detail in a future paper. By design, our sample of dormant galaxies are among the sources with the highest Balmer break ratios and lowest H$\alpha$ EWs, populating the tail of the distribution.

\subsection{Star formation histories}

To confirm that our galaxies are indeed in a dormant phase, we analyzed the SFHs of the sample derived from \texttt{Bagpipes}. Fig. \ref{fig:sfhs} shows the specific SFHs (sSFHs) of the 14 dormant galaxies selected in this work. We find that all of them share a similar SFH, as they all halted their star formation between 10 and 25 Myr before observation, confirming that these are currently galaxies in an off mode. 
By comparing the mean sSFH of the dormant galaxies against that of the entire DJA sample, we find a significant contrast, as the ``average'' galaxy is increasing star formation, although we cannot draw conclusions, as ours is not a complete sample and is likely biased toward galaxies with larger star formation. However, this clearly shows that dormant galaxies are distinct from average galaxies.
There is one galaxy that clearly shows lower specific SFR (sSFR) values at all age bins compared to the rest: the galaxy with ID 4188, which is the most magnified galaxy in our sample due to lensing ($\mu=35_{-15}^{+68}$; \citealt{Meena23}). The great magnification of this source explains that we are capable of obtaining its spectrum with high S/N despite of its low sSFR. The SFH of this galaxy is also unlike the others as, while the rest of the sample experiences a rise in SFR before it shuts off, this galaxy started slowly decreasing its star formation 250 Myr before observation and has continued ever since, indicating that the mechanism that halted star formation in this object may have been different from the others. The off-mode timescales of these galaxies (at least $10-25$ Myr) are consistent with expectations from simulations in \citet{Dome24}, which find from models that $4<z<8$ galaxies in the mass range $M*=10^7-10^9M_\odot$ have dormant phases that typically last $\sim20-40$ Myr, before they start a new phase of star formation. This suggests that the galaxy burstiness could be driven by the inflow and outflow of gas internally within the galaxy, as otherwise we would encounter longer periods in the off mode \citep{McClymont25}. However, as we cannot know for how much longer they will stay in this off mode, it is possible that other mechanisms are acting on these galaxies over longer timescales, such as AGN feedback or environmental quenching, which can halt star formation for $>100$ Myr. This is especially relevant for the one galaxy in our sample that decreased star formation 250 Myr before observation, although it was not a sudden stop. 

To characterize the recent downturn in star-formation more in detail, we computed the $\textrm{SFR}_{10}/\textrm{SFR}_{100}$ (that is, the ratio between the SFR averaged over the last 10 Myr and that of the last 100 Myr) from \texttt{Bagpipes}, which is shown in Table \ref{tab:dormants}. We observe that these ratios ratio is always below 20\%, with a median of 1.3\%; nevertheless, their uncertainties are larger than the values themselves, resulting in lower limits of $\textrm{SFR}_{10}/\textrm{SFR}_{100}\sim0$, which is equivalent to the $\textrm{SFR}_{3}/\textrm{SFR}_{50}\sim0$
criterion that \citet{Gelli25} used to select dormant galaxies from the \texttt{SERRA} simulation. However, three of our sources do not fulfill this criterion, as their lower limits are $>1\%$, and thus, they would be considered galaxies in a downturn phase of their SFH, according to their selection. 

The mass-weighted ages of this sample of dormant galaxies range from $\sim65-400$ Myr, with the median being 200 Myr. Thus, the bulk of these galaxies was formed about $\sim200$ Myr before the time of observation, which corresponds to $z\sim6-8.3$, depending on the galaxy. Although we cannot predict the future SFH of these galaxies, it is possible that they resume star formation within the following $\sim15-30$ Myr, as is expected from a bursty SFH; nevertheless, we cannot rule out that some of these galaxies will remain quiescent long-term, especially the most massive ones, which are less likely to present a bursty SFH. It is thus possible that we are catching some of these galaxies right on their way to long-term quiescence. It will be interesting to compare the number densities of the dormant population with the build-up of the quiescent galaxy population over the same redshift range to quantify this in more detail in the future.

\subsection{Populating the low-SFR end of the main sequence}
\begin{figure}
\includegraphics[width=\columnwidth]{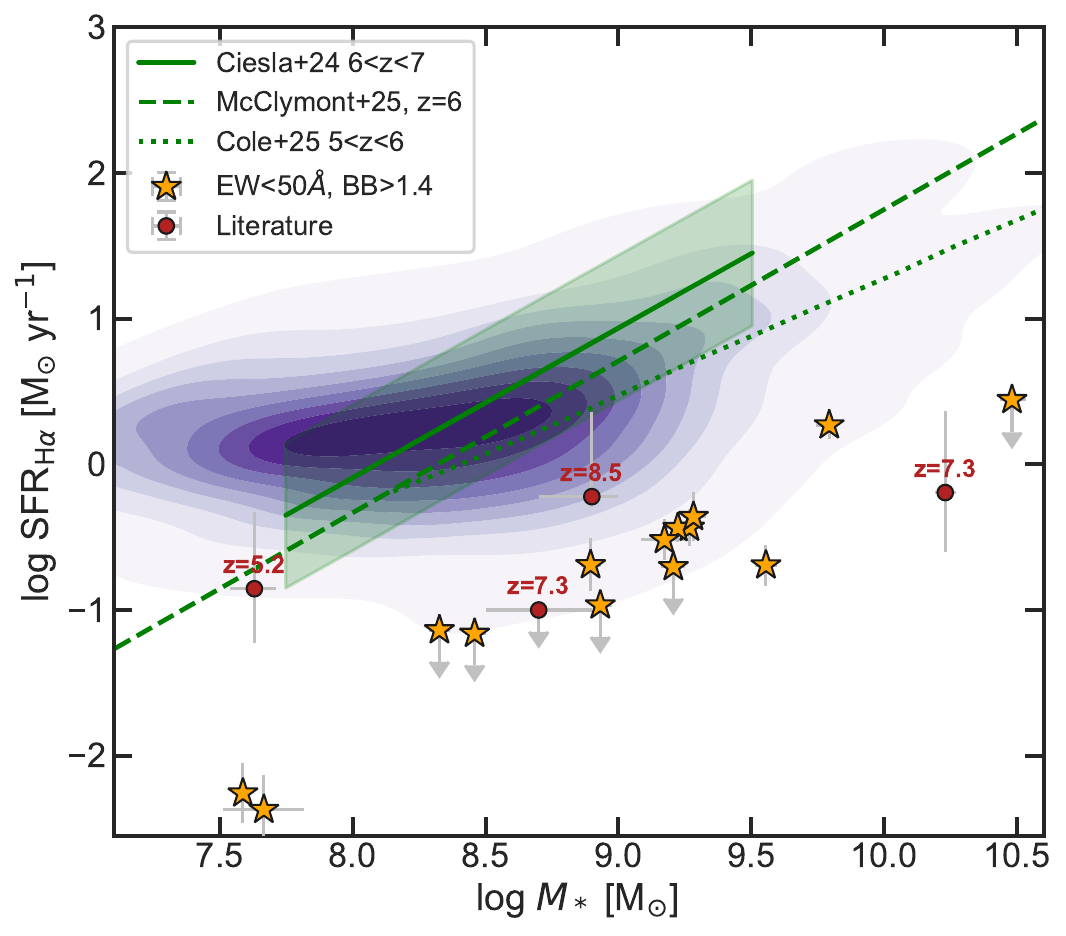}
    \caption{Distribution of mass and H$\alpha$-derived SFR of our sample of 1,598 galaxies (purple contours). Also shown are the main sequence fits from \citet{Ciesla24} for $6<z<7$ galaxies, from \citet{McClymont25} for $z\sim6$, and from \citet{Cole25} for $5<z<6$. The shaded green area represents the $0.5$ dex scatter in the main sequence reported by \citet{Ciesla24}. The starred points are the 14 dormant galaxies selected in this work and the filled circles are those reported to date at $z>5$: at $z=5.2$ \citep{Strait+23}, at $z=7.3$ \citep{Looser+24}, at $z=7.3$ \citep{Weibel+24}, and at $z=8.5$ \citep{Baker+25}; H$\alpha$ cannot be observed for the sources at $z>7.4$, and therefore we plot SFR$_{10}$ instead of SFR$_{\textrm{H}\alpha}$ for those literature sources. We observe that all of our dormant galaxies lie below the main sequence and span a stellar mass range of $M_*\sim10^{7.6}-10^{10.5}M_\odot$, without an empty gap between the low- and high-mass regime.}
    \label{fig:main_sequence}
\end{figure}

The width of the star-forming main sequence of galaxies provides important information on the burstiness of galaxies' SFHs, especially when comparing the widths as measured using different SFR indicators such as H$\alpha$ and UV luminosities \citep[e.g.,][]{Pirie24}. It is therefore interesting to compare our dormant galaxies to the rest of the population and place them in the main sequence. Fig. \ref{fig:main_sequence} shows the position of our selected dormant galaxies in the star-forming main sequence, where the SFRs were derived from the H$\alpha$ fluxes following equations \ref{eq:att} and \ref{eq:SFR}. The SFR values derived from H$\alpha$ agree with the SFR$_{10}$ obtained from SED fitting within their uncertainties, although H$\alpha$ SFRs are only upper limits and the uncertainties in SFR$_{10}$ for our galaxy selection reach 1 dex due to the low star formation values. As ours is not a complete sample, we do not aim to compute the main sequence from our catalog. Instead, we show where our galaxies lie in comparison with \citet{Ciesla24} and \citet{Cole25}, based on photometric observations at $6<z<7$ and $5<z<6$, respectively; and \citet{McClymont25}, based on simulations. We find that all the dormant galaxies in our sample lie below the main sequence, which has a scatter of $\sim0.5$ dex according to the findings in \citet{Ciesla24}. 

We also find that our selection of dormant galaxies spans a stellar mass range of $M_*\sim10^{7.6}-10^{10.5}M_\odot$, indicating that galaxies in an off-state are present across all probed mass scales above a detection threshold. This is in contrast with a bimodal scenario that could be extrapolated from the previously reported literature values (also shown in Fig. \ref{fig:main_sequence}) -- with a distinct population of low-mass mini-quenched galaxies with $M_*<10^9M_\odot$ \citep[e.g.,][]{Strait+23,Looser+24,Baker+25} and a separate population of massive quiescent systems with $M_*>10^{10}M_\odot$ \citep[e.g.,][]{Weibel+24,Baker+25b}. 
Our findings reveal the existence of an intermediate-mass population: this suggests a more continuous distribution of dormant galaxies in stellar mass, without an empty gap between low- and high-mass regimes, although we cannot rule out a bimodality in the mass distribution due to the incompleteness of our sample. 
The existence of these intermediate-mass dormant galaxies shows that bursty SFHs are found at all stellar masses and are not exclusive to low-mass systems. 
Nevertheless, the most massive galaxies in this sample might be the progenitors of massive quiescent galaxies at lower redshifts.

\section{Summary and conclusions}
\label{sec:conclu}

In this work we performed a systematic search for dormant galaxies that halted star formation at least 10 Myr before the time of observation at $5<z<7.4$. To do this, we made use of the entirety of the NIRSpec prism data available in the DJA and performed an empirical selection based on low rest-frame H$\alpha$ EWs and strong Balmer breaks. We identified 14 dormant galaxies that halted star formation between 10 and 25 Myr before the time of observation; they all lie below the main sequence and have stellar masses ranging from $M_*\sim10^{7.6}-10^{10.5}M_\odot$. This demonstrates that dormant galaxies are distributed over a large range of stellar masses and are compatible with temporary quenching within bursty SFHs. 
Nevertheless, in this work we only selected the most extreme sources (i.e., those with the lowest EWs and the strongest Balmer breaks), in order to reliably identify the off mode of star formation; there is a significant number of galaxies below the main sequence that could be on their way to a short dormant phase. 

These results show that only $\sim1\%$ of the galaxies in the DJA at $5<z<7.4$ are dormant, and they can be selected through spectroscopic features. Nevertheless, without a complete sample of galaxies across the main sequence, it is not possible to quantify the fraction of dormant galaxies within the overall galaxy population. In this regard, future surveys targeting complete samples using NIRSpec prism data will be fundamental to quantifying this distribution. Moreover, most targeted galaxies present in the DJA are above the main sequence and on the high-mass end; performing such observations on galaxies below the main sequence would allow us to  identify a larger number of dormant galaxies and more properly characterize the timescales of burstiness and the distribution of times that distant galaxies spend in the off mode.

The H$\alpha$ emission line is only covered by NIRSpec up to $z\lesssim7.4$, and thus, with our method, we could only find dormant galaxies up to that limit. Nevertheless, using bluer emission lines such as the [OIII] doublet combined with H$\beta$ in a similar approach could enable searches for dormant galaxies out to higher redshifts, both in the archive and in future observations. The combination of the different selection techniques and the new, complete samples that will soon be observed with the JWST NIRSpec prism will reveal the population and nature of these dormant objects.

\begin{acknowledgements}
The authors thank Tobias Looser and William Baker for fruitful discussions and guidance, and Lukas Furtak and Adi Zitrin for providing the lensing models for UNCOVER. This work is based on observations made with the NASA/ESA/CSA James Webb Space Telescope. The raw data were obtained from the Mikulski Archive for Space Telescopes at the Space Telescope Science Institute, which is operated by the Association of Universities for Research in Astronomy, Inc., under NASA contract NAS 5-03127 for \textit{JWST}. Some of the data products presented herein were retrieved from the Dawn JWST Archive (DJA). DJA is an initiative of the Cosmic Dawn Center, which is funded by the Danish National Research Foundation under grant No. 140 (DNRF140). This work has received funding from the Swiss State Secretariat for Education, Research and Innovation (SERI) under contract number MB22.00072, as well as from the Swiss National Science Foundation (SNSF) through project grant 200020\_207349.
\end{acknowledgements}

\bibliographystyle{aa}
\bibliography{references}

\begin{appendix}
\onecolumn
\section{Spectra}
\begin{figure*}
\includegraphics[width=\linewidth]{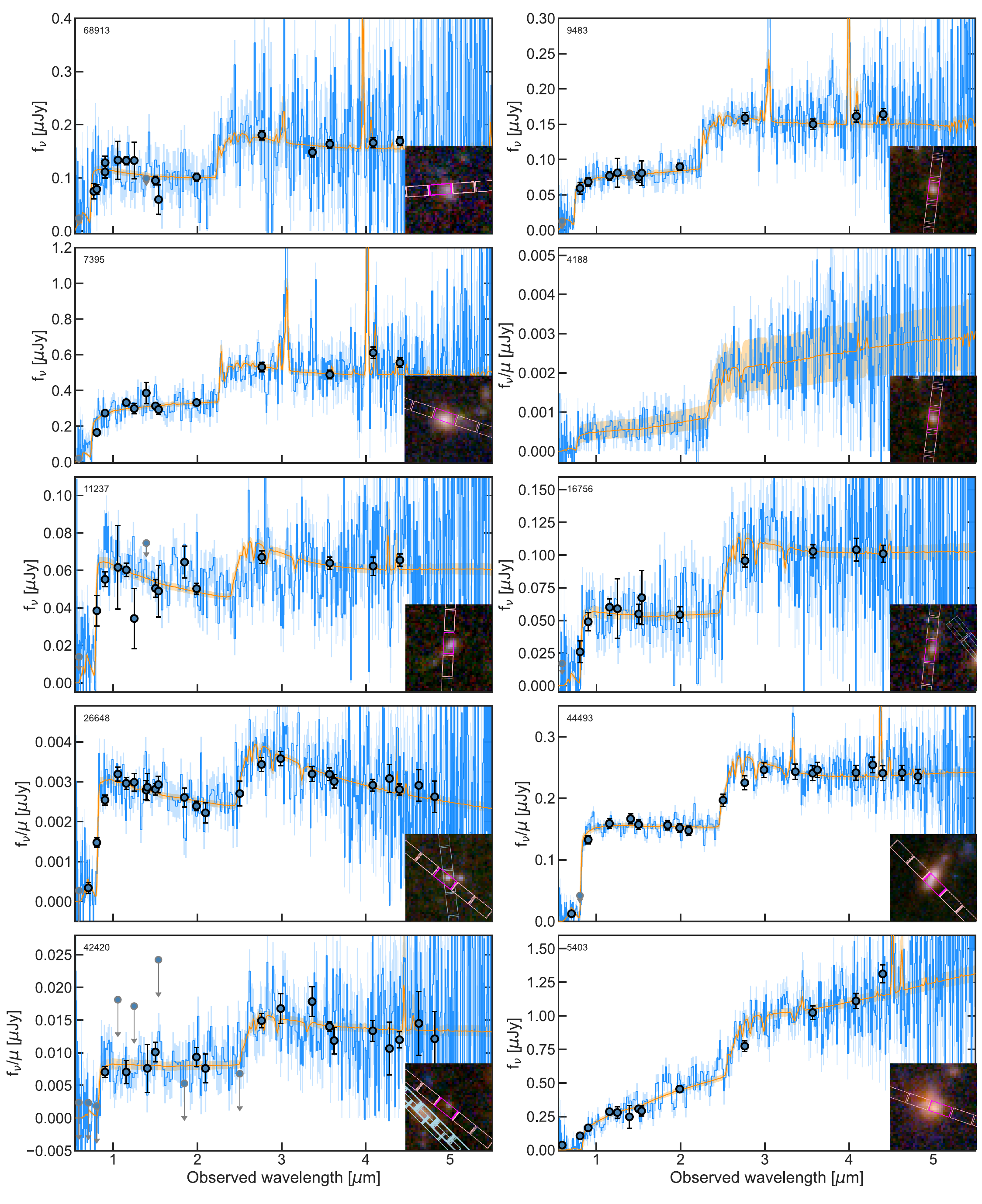}
    \caption{Observed $f_\nu$ spectra (blue) and posterior spectra from \texttt{Bagpipes} (orange) for four of the dormant galaxies selected in this work. The filled circles are the photometry values, which are used to calibrate the spectra. We observe strong Balmer breaks and weak emission lines. Inset are the RGB cutouts of the galaxies, with the slits used for these observations overplotted in pink.} 
    \label{fig:spectra}
\end{figure*}

In Fig. \ref{fig:spectra} we show the spectra of the full sample of dormant galaxies that we selected in this work.

\begin{figure*}
\includegraphics[width=\linewidth]{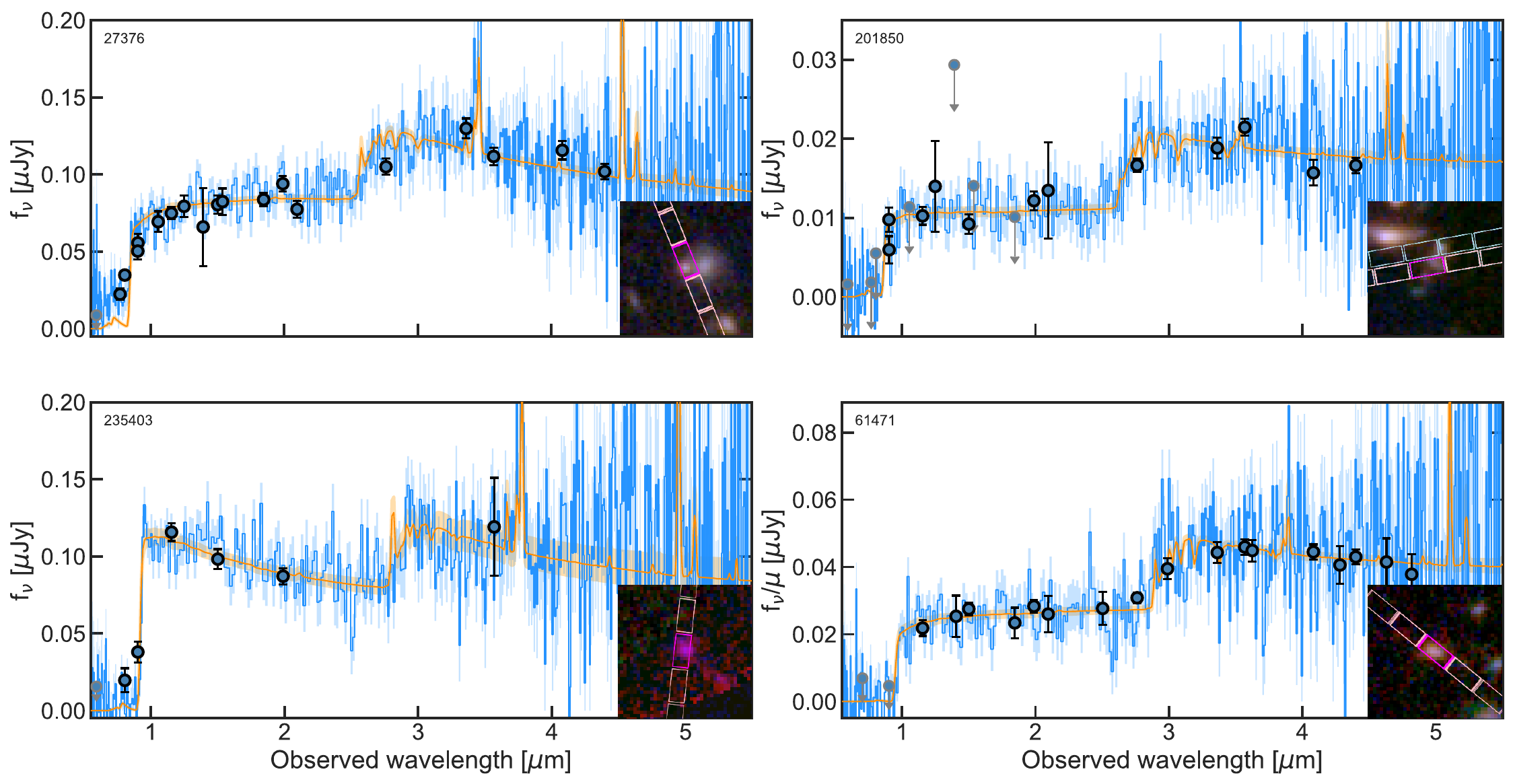}
\captionsetup{labelformat=empty,justification=raggedright,singlelinecheck=false}
\caption{Fig.~\ref{fig:spectra}. continued.}
\end{figure*}
\end{appendix}

\end{document}